\documentclass[pdflatex,sn-mathphys-num]{sn-jnl}%
\usepackage{graphicx}%
\usepackage{multirow}%
\usepackage{amsmath,amssymb,amsfonts}%
\usepackage{amsthm}%
\usepackage{mathrsfs}%
\usepackage[title]{appendix}%
\usepackage{xcolor}%
\usepackage{textcomp}%
\usepackage{manyfoot}%
\usepackage{booktabs}%
\usepackage{graphicx}   
\usepackage{cleveref}

\usepackage{subcaption}

\usepackage{algorithmicx}%
\usepackage{algpseudocode}%
\usepackage{listings}%
\usepackage[ruled,vlined]{algorithm2e}
\usepackage{tikz}
\usetikzlibrary{shapes.geometric, arrows, positioning}

\begin{document}

\title{Colormap-Enhanced Vision Transformers for MRI-Based Multiclass (4-Class) Alzheimer’s Disease Classification}

\author*[1]{\fnm{Faisal Ahmed} }\email{ahmedf9@erau.edu}

\affil*[1]{\orgdiv{Department of Data Science and Mathematics}, \orgname{Embry-Riddle Aeronautical University}, \orgaddress{\street{3700 Willow Creek Rd}, \city{Prescott}, \postcode{86301}, \state{Arizona}, \country{USA}}}

\abstract{

Magnetic Resonance Imaging (MRI) plays a pivotal role in the early diagnosis and monitoring of Alzheimer’s disease (AD). However, the subtle structural variations in brain MRI scans often pose challenges for conventional deep learning models to extract discriminative features effectively. In this work, we propose \textbf{PseudoColorViT-Alz}, a colormap-enhanced Vision Transformer framework designed to leverage pseudo-color representations of MRI images for improved Alzheimer’s disease classification. By combining colormap transformations with the global feature learning capabilities of Vision Transformers, our method amplifies anatomical texture and contrast cues that are otherwise subdued in standard grayscale MRI scans.  

We evaluate \textbf{PseudoColorViT-Alz} on the OASIS-1 dataset using a four-class classification setup (non-demented, moderate dementia, mild dementia, and very mild dementia). Our model achieves a state-of-the-art accuracy of \textbf{99.79\%} with an AUC of \textbf{100\%}, surpassing the performance of recent 2024–2025 methods, including CNN-based and Siamese-network approaches, which reported accuracies ranging from 96.1\% to 99.68\%.  These results demonstrate that pseudo-color augmentation combined with Vision Transformers can significantly enhance MRI-based Alzheimer’s disease classification. \textbf{PseudoColorViT-Alz} offers a robust and interpretable framework that outperforms current methods, providing a promising tool to support clinical decision-making and early detection of Alzheimer’s disease.

}




\maketitle

\section{Introduction}\label{sec1}

Magnetic Resonance Imaging (MRI) is a widely used, non-invasive tool for assessing brain structure and detecting neurodegenerative diseases such as Alzheimer’s disease (AD)~\cite{jack2008alzheimer,mosconi2005early}. Despite its clinical importance, automated analysis of MRI scans remains challenging due to the limited availability of annotated datasets, inter-patient variability, and subtle structural changes associated with early-stage AD, which are difficult to detect using conventional methods~\cite{liu2021deep}.

Recent advancements in deep learning, particularly Convolutional Neural Networks (CNNs), have demonstrated promising results in automated disease classification from brain MRI images~\cite{zhang2021multi}. However, CNNs typically focus on local patterns and often struggle to capture long-range spatial relationships that are crucial for detecting subtle neurodegenerative changes~\cite{ahmed2023topo}. Moreover, most state-of-the-art models require large annotated datasets, limiting their applicability in scenarios where medical imaging data is scarce~\cite{hernandez2019ensemble}.

Vision Transformers (ViTs) have recently emerged as powerful models for image classification, effectively capturing long-range dependencies and global contextual information~\cite{dosovitskiy2021image,liu2021swin}. However, the direct application of ViTs to grayscale MRI scans poses challenges, as most ViTs are pretrained on large-scale RGB natural image datasets. Existing adaptations, such as replicating the grayscale channel to create pseudo-RGB inputs or training ViTs from scratch, often fail to fully exploit structural and textural information in MRI scans, leading to suboptimal performance under limited data scenarios~\cite{tougaccar2020deep}.

To address these challenges, we propose \textbf{PseudoColorViT-Alz}, a colormap-enhanced Vision Transformer framework that transforms grayscale MRI scans into pseudo-color representations, enabling better utilization of pretrained ViTs. This approach enhances contrast and structural cues in MRI images, allowing the model to capture subtle anatomical changes associated with Alzheimer’s disease (AD) without requiring extensive retraining.

We evaluate \textbf{PseudoColorViT-Alz} on the OASIS-1 dataset using a four-class classification setup (non-demented, moderate dementia, mild dementia, and very mild dementia). Our method achieves a state-of-the-art accuracy of 99.79\% with an AUC of 100\%, surpassing recent 2024–2025 CNN-based and Transformer approaches, which report accuracies between 96.1\% and 99.68\%. These results demonstrate that pseudo-color augmented Vision Transformers offer a robust and interpretable framework for MRI-based Alzheimer’s disease classification, providing an effective tool for early detection and clinical decision support.

\begin{figure}[t!]
	\centering
	\subfloat[\scriptsize Non-demented MRI sample.\label{fig:ND}]{%
		\includegraphics[width=0.2\linewidth]{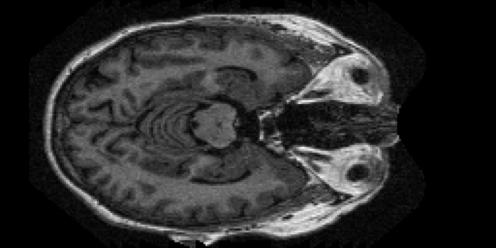}}
	\hfill
	\subfloat[\scriptsize Moderate dementia MRI sample.\label{fig:MD}]{%
		\includegraphics[width=0.2\linewidth]{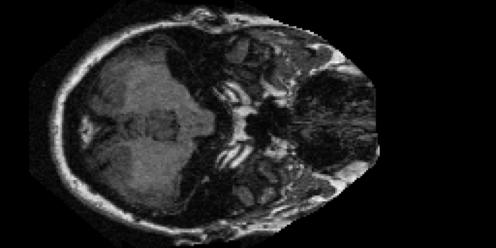}}
	\hfill
	\subfloat[\scriptsize Mild dementia MRI sample.\label{fig:MiD}]{%
		\includegraphics[width=0.2\linewidth]{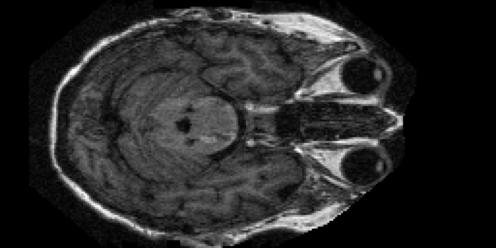}}
    \hfill
	\subfloat[\scriptsize Very mild dementia MRI sample.\label{fig:VMD}]{%
		\includegraphics[width=0.2\linewidth]{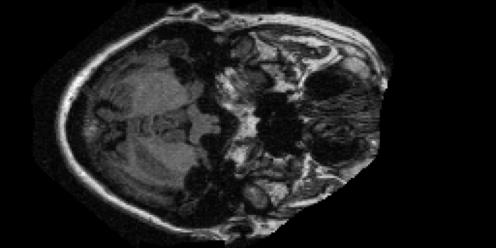}}

	\caption{\footnotesize Representative brain MRI samples from the OASIS-1 dataset illustrating the four Alzheimer’s disease categories used in this study.}
	\label{fig:image-samples}
\end{figure}

\section{Related Works}\label{sec2}

Automated analysis of brain MRI scans has become central to diagnosing neurodegenerative diseases such as Alzheimer’s disease (AD). Early machine learning methods relied heavily on handcrafted features, voxel-based morphometry, and statistical models. For example, support vector machines (SVMs) and classical CNN-based approaches~\cite{Kloppel2008SVM, Zhang2022VBMCNN} demonstrated initial success in detecting structural changes associated with AD but were limited in feature generalization and scalability across diverse datasets.

With the advent of deep learning, Convolutional Neural Networks (CNNs) became the dominant paradigm for MRI-based AD classification. Models such as 3D-CNN~\cite{ebrahimi2021convolutional}, deep learning-based ensemble method ~\cite{fathi2024deep}, and DenseNet architectures~\cite{wang2021densecnn} achieved significant improvements in accuracy by learning hierarchical feature representations from MRI volumes. Ensemble-based CNNs further improved performance by combining multiple architectures~\cite{fathi2024deep}. However, these methods often require large annotated datasets and are sensitive to class imbalance, which can limit their generalization to smaller cohorts or different imaging centers.

Recent studies have explored hybrid and topological techniques to enhance feature extraction. Bazin and Pham~\cite{bazin2007topology} proposed a topology-preserving, anatomy-driven brain MRI segmentation framework (TOADS) that enforces structural and topological consistency during tissue classification. Feature selection methods such as mRMR~\cite{alshamlan2024improving, alshamlan2023identifying} have also been applied to reduce irrelevant features and improve interpretability in MRI-based AD diagnosis. While these approaches advance diagnostic accuracy, they still face limitations in computational efficiency and robustness, particularly on moderately sized datasets. Furthermore, a growing body of research has demonstrated the effectiveness of topological data analysis (TDA) for medical image classification. By capturing intrinsic geometric and structural properties of data, TDA-based approaches have been successfully applied across a range of medical imaging tasks, yielding improved robustness and discriminative performance~\cite{ahmed2025topo, ahmed2023tofi, ahmed2023topological, ahmed2023topo, yadav2023histopathological, ahmed2025topological, ahmed20253d}.

Vision Transformers (ViTs)~\cite{dosovitskiy2021image,liu2021swin} have recently emerged as a promising alternative to CNNs, leveraging global self-attention mechanisms to capture long-range dependencies effectively. ViTs have shown potential in medical imaging tasks, including brain MRI analysis~\cite{sankari2025hierarchical, dhinagar2023efficiently}, by providing superior global feature representation. Nonetheless, their direct application to grayscale MRI scans is challenging due to high computational cost and the need for large-scale training data. Standard adaptations such as channel replication or training from scratch often fail to exploit structural and textural information optimally. More applications of transfer learning and Vision Transformers in medical image analysis are explored in the following studies:~\cite{ahmed2025hog, ahmed2025ocuvit, ahmed2025robust, ahmed2025histovit, ahmed2025transfer, ahmed2025repvit, ahmed2025pseudocolorvit, rawat2025efficient}.

To overcome these limitations, we propose \textbf{PseudoColorViT-Alz}, a colormap-enhanced Vision Transformer framework that transforms grayscale MRI scans into pseudo-color representations. This approach enhances subtle anatomical and structural patterns, allowing the model to fully leverage the global contextual modeling capability of Transformers. Our method achieves state-of-the-art performance on OASIS-1 for four-class Alzheimer’s disease classification, demonstrating robustness even with moderately sized datasets.

\begin{figure*}[t!]
    \centering
    \includegraphics[width=\linewidth]{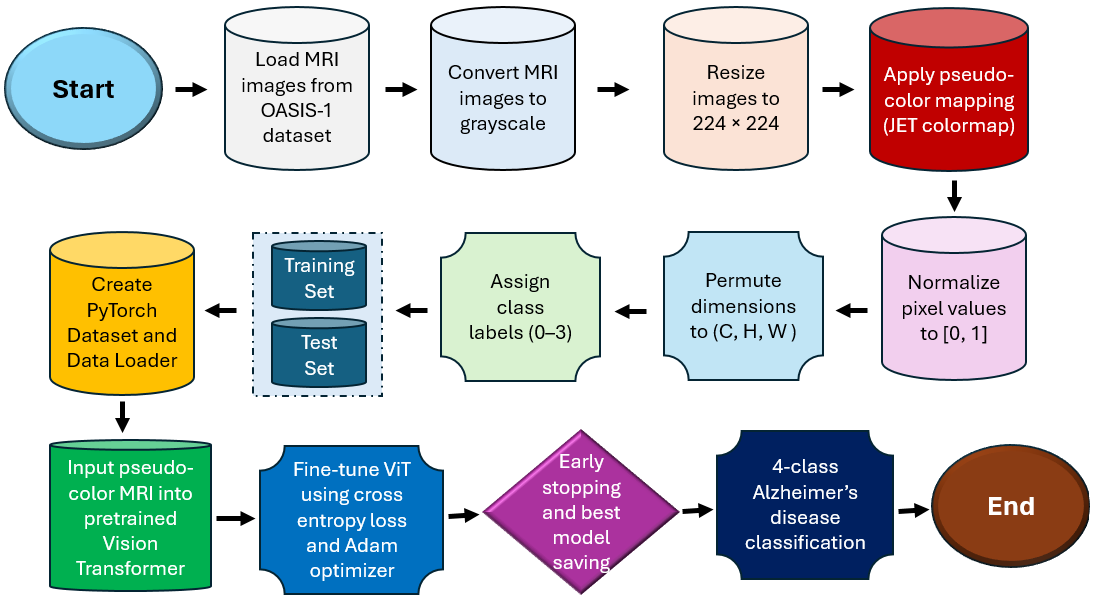}
    \caption{\footnotesize \textbf{PseudoColorViT-Alz preprocessing pipeline.} Overview of the complete data preprocessing workflow, illustrating the steps from raw brain MRI image acquisition, grayscale conversion, and pseudo-color enhancement to normalization, dataset construction, and input preparation for Vision Transformer–based multiclass Alzheimer’s disease classification.}
    \label{fig:Preprocess}
\end{figure*}

\section{Methodology}\label{sec:method}

\begin{figure*}[t!]
    \centering
    \includegraphics[width=\linewidth]{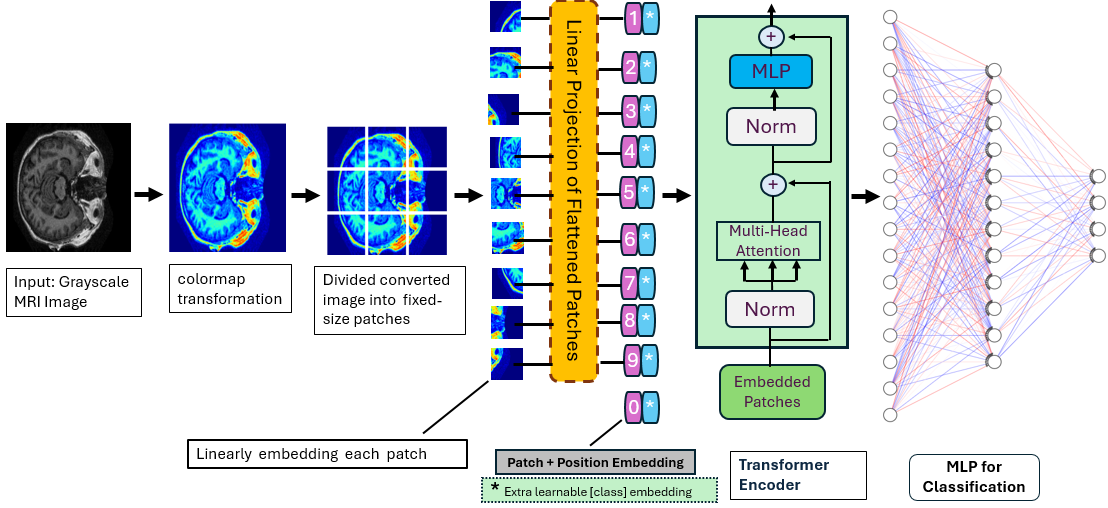}
    \caption{\footnotesize \textbf{PseudoColorViT pipeline.} The proposed framework follows the Vision Transformer architecture introduced in~\cite{dosovitskiy2021image}. A 2D grayscale medical image is first converted into a pseudo-color representation, yielding a three-channel input that preserves anatomical structure while enhancing texture and contrast information. The pseudo-colored image is then partitioned into fixed-size patches, which are linearly projected and augmented with positional encodings. A learnable classification token is appended to the patch sequence and processed by the Transformer encoder. The final encoded representation is passed through a classification head to produce the disease prediction.}
    \label{fig:flowchart}
\end{figure*}

This section describes the proposed methodology for multiclass Alzheimer’s disease classification using brain MRI images from the OASIS-1 dataset. The overall pipeline consists of MRI image preprocessing with pseudo-color enhancement, dataset construction, Vision Transformer fine-tuning, model training with early stopping, and comprehensive evaluation using classification and ROC-based metrics.

\subsection{Data Preparation and Pseudo-Color Preprocessing}

Brain MRI scans from four Alzheimer’s disease categories—non-demented, moderate dementia, mild dementia, and very mild dementia—are first converted to grayscale intensity images and resized to a fixed spatial resolution of $224 \times 224$ pixels. Since Vision Transformers are pretrained on RGB natural images, grayscale MRI scans are transformed into pseudo-color representations using the \textit{jet} colormap. This process enhances subtle intensity variations and anatomical structures relevant to neurodegenerative progression.

Let $\mathbf{I}_{\text{gray}} \in \mathbb{R}^{224 \times 224}$ denote a grayscale MRI image. Pseudo-color mapping is applied as
\[
\mathbf{I}_{\text{rgb}} = \text{Colormap}\left(\frac{\mathbf{I}_{\text{gray}}}{255}\right),
\]
where $\mathbf{I}_{\text{rgb}} \in \mathbb{R}^{224 \times 224 \times 3}$ represents the three-channel pseudo-color image. Pixel values are normalized to the range $[0,1]$ to stabilize training:
\[
\mathbf{I}_{\text{norm}} = \frac{\mathbf{I}_{\text{rgb}}}{255}.
\]

Each image tensor is permuted to PyTorch’s channel-first format $(C, H, W)$ before being passed to the network.

\subsection{Dataset Construction and Splitting}

The final dataset is constructed by concatenating pseudo-color MRI images from all four classes and assigning integer labels as follows: $0$ (non-demented), $1$ (mild dementia), $2$ (moderate dementia), and $3$ (very mild dementia). The combined dataset is randomly split into training and testing subsets using an 80:20 ratio. A custom PyTorch \texttt{Dataset} class is implemented to map each image $\mathbf{I}_i$ to its corresponding label $y_i \in \{0,1,2,3\}$. Mini-batches of size 32 are generated using \texttt{DataLoader} with shuffling enabled for training.

\subsection{Vision Transformer Architecture}

The backbone model employed is the pretrained \texttt{google/vit-base-patch16-224} Vision Transformer, fine-tuned for four-class Alzheimer’s disease classification. Each input image $\mathbf{I} \in \mathbb{R}^{3 \times 224 \times 224}$ is divided into non-overlapping patches of size $16 \times 16$, yielding $N=196$ patches. Each patch is flattened and linearly projected into a latent embedding space:
\[
\mathbf{E}_i = \mathbf{W}\cdot \text{Flatten}(\text{Patch}_i) + \mathbf{b}, \quad i=1,\dots,N.
\]

A learnable classification token $\mathbf{x}_{\text{cls}}$ is prepended, and positional embeddings $\mathbf{p}_i$ are added to preserve spatial information:
\[
\mathbf{z}^0 = [\mathbf{x}_{\text{cls}}, \mathbf{E}_1 + \mathbf{p}_1, \dots, \mathbf{E}_N + \mathbf{p}_N].
\]

The sequence is processed by a stack of transformer encoder layers, each comprising multi-head self-attention and feed-forward networks. The self-attention mechanism is defined as
\[
\text{Attention}(\mathbf{Q}, \mathbf{K}, \mathbf{V}) =
\text{Softmax}\left(\frac{\mathbf{QK}^\top}{\sqrt{d_k}}\right)\mathbf{V},
\]
followed by residual connections and layer normalization:
\[
\mathbf{z}^{\ell+1} = \text{LayerNorm}(\mathbf{z}^{\ell} + \text{FFN}(\mathbf{z}^{\ell})).
\]

The final hidden state of the classification token $\mathbf{z}^L_{\text{cls}}$ is passed to a fully connected classification head with Softmax activation:
\[
\hat{\mathbf{y}} = \text{Softmax}(\mathbf{W}_{\text{cls}}\mathbf{z}^L_{\text{cls}} + \mathbf{b}_{\text{cls}}).
\]

\subsection{Training Strategy}

Model optimization is performed using the categorical cross-entropy loss:
\[
\mathcal{L} = -\frac{1}{B}\sum_{i=1}^{B}\sum_{c=1}^{4} y_{i,c}\log(\hat{y}_{i,c}),
\]
where $B$ denotes the batch size and $C=4$ the number of classes. The Adam optimizer is used with a learning rate of $1\times10^{-4}$. Training is conducted for a maximum of 50 epochs with early stopping (patience = 2) based on test accuracy. The best-performing model is saved automatically.

\subsection{Evaluation Metrics}

\begin{figure*}[t!]
    \centering
    \subfloat[\scriptsize One-vs-rest ROC curves with AUC scores.\label{fig:auc}]{
        \includegraphics[width=0.45\linewidth]{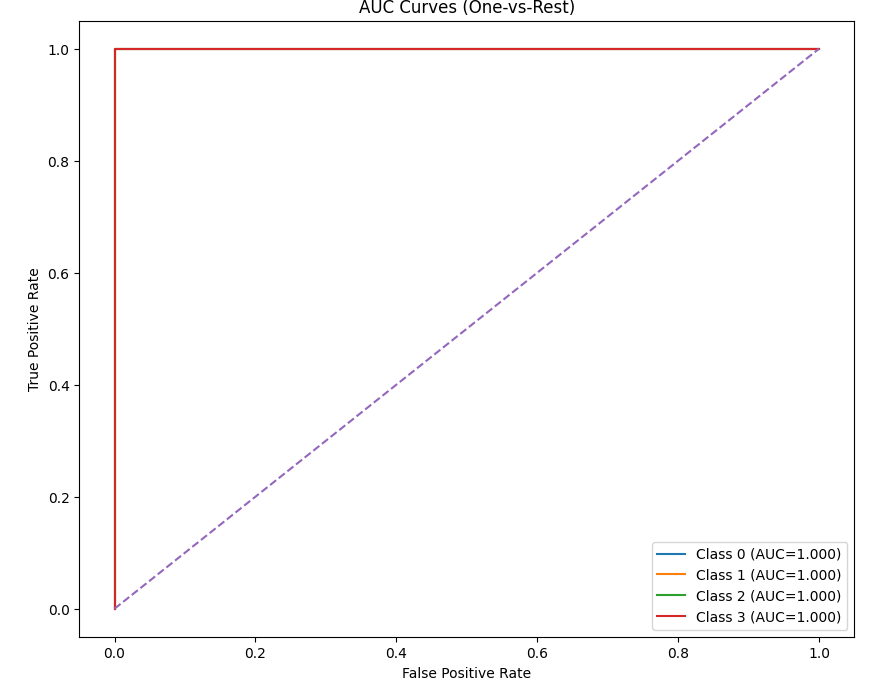}}
    \hfill
    \subfloat[\scriptsize Confusion matrix for four-class Alzheimer’s disease classification.\label{fig:conf}]{
        \includegraphics[width=0.45\linewidth]{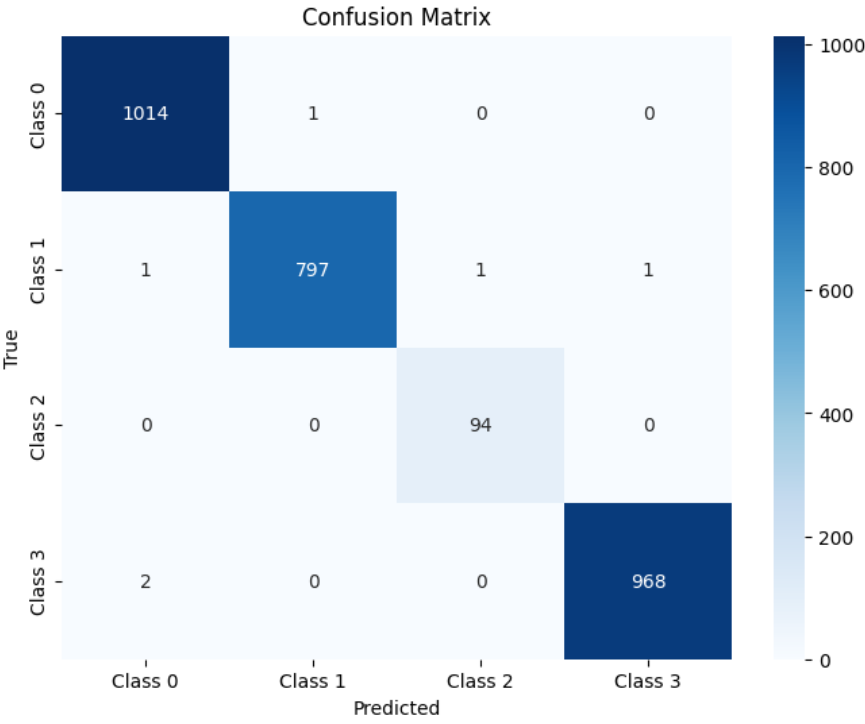}}
    
    \caption{\footnotesize Performance evaluation of \textbf{PseudoColorViT-Alz} on the OASIS-1 dataset. The left panel shows one-vs-rest ROC curves and corresponding AUC values for each Alzheimer’s disease class, while the right panel presents the confusion matrix illustrating class-wise prediction performance.}
    \label{fig:auc_conf}
\end{figure*}

Model performance is assessed on the held-out test set using accuracy, macro-averaged precision, recall, and multi-class area under the ROC curve (AUC) using a one-vs-rest strategy. Final predictions are obtained as
\[
\hat{y}_i = \arg\max_c \hat{y}_{i,c}.
\]

Additionally, confusion matrices are generated to visualize class-wise prediction behavior \Cref{fig:conf}, and ROC curves \Cref{fig:auc} are plotted for each Alzheimer’s disease category to analyze sensitivity–specificity trade-offs. All experiments are implemented in PyTorch with GPU acceleration when available.

\begin{algorithm}[H]
\SetAlgoNlRelativeSize{0}
\DontPrintSemicolon
\caption{PseudoColorViT-Alz: MRI-Based Multiclass Alzheimer’s Disease Classification}
\label{alg:pcvit_alz}

\KwIn{MRI dataset $\mathcal{D}=\{(\mathbf{I}_i, y_i)\}_{i=1}^{N}$, pretrained ViT model, number of epochs $E$, patience $p$, number of classes $C=4$}
\KwOut{Best trained Vision Transformer model and evaluation metrics}

\textbf{Preprocessing and Pseudo-Color Enhancement:} \\
\For{$i \gets 1$ \KwTo $N$}{
    Convert MRI image $\mathbf{I}_i$ to grayscale\;
    Resize $\mathbf{I}_i$ to $(224 \times 224)$\;
    Apply pseudo-color mapping using a colormap to obtain RGB representation\;
    Normalize pixel intensities: $\mathbf{I}_i \gets \mathbf{I}_i / 255$\;
    Permute tensor dimensions to $(C,H,W)$\;
}

Split dataset $\mathcal{D}$ into training and testing sets using an 80:20 ratio\;

\textbf{Model Initialization:} \\
Load pretrained \texttt{ViT-base-patch16-224}\;
Replace classification head with a $C$-class Softmax layer\;
Initialize Adam optimizer with learning rate $\eta = 1\times10^{-4}$\;
Initialize categorical cross-entropy loss function\;

\textbf{Training and Validation Loop:} \\
Initialize best test accuracy $\alpha^{*} \gets 0$ and counter $\delta \gets 0$\;

\For{$\text{epoch} \gets 1$ \KwTo $E$}{
    Set model to \texttt{train} mode\;
    \ForEach{mini-batch $(\mathbf{X}, \mathbf{y}) \in \mathcal{D}_{train}$}{
        Forward pass: $\hat{\mathbf{y}} \gets \text{ViT}(\mathbf{X})$\;
        Compute loss: $\mathcal{L} \gets \text{CrossEntropy}(\hat{\mathbf{y}}, \mathbf{y})$\;
        Backpropagate gradients and update model parameters\;
    }
    Compute and log training loss and accuracy\;

    Set model to \texttt{eval} mode\;
    \ForEach{mini-batch $(\mathbf{X}, \mathbf{y}) \in \mathcal{D}_{test}$}{
        Forward pass without gradient computation\;
        Accumulate test loss and accuracy\;
    }
    Log validation results to CSV file\;

    \textbf{Early Stopping Criterion:} \\
    \If{current test accuracy $> \alpha^{*}$}{
        Save model weights as \texttt{best\_model.pth}\;
        Update $\alpha^{*}$\;
        Reset counter $\delta \gets 0$\;
    }
    \Else{
        Increment counter $\delta \gets \delta + 1$\;
    }
    \If{$\delta \geq p$}{
        Terminate training loop\;
    }
}

\textbf{Final Evaluation:} \\
Load best saved model parameters\;
Generate predictions using Softmax probabilities\;
Compute accuracy, macro-averaged precision, recall, and multi-class ROC-AUC (OvR)\;
Generate confusion matrix and class-wise ROC curves\;

\Return{Optimized PseudoColorViT-Alz model and performance metrics}
\end{algorithm}

\section{Experiment}

\subsection{Datasets}


\noindent {\bf OASIS-1 dataset} \cite{Marcus2007OpenAccess}

The OASIS-1 dataset is a widely used cross-sectional structural MRI collection containing T1-weighted brain scans from 416 adult subjects aged 18–96 years. Each participant contributes three or four individual T1-weighted MRI scans acquired within the same session, enabling high signal-to-noise averaging and robust morphometric analysis. Among the 416 subjects, 100 individuals over the age of 60 are clinically diagnosed with very-mild to moderate Alzheimer’s disease (AD), while the remaining 316 subjects serve as nondemented controls. Because each subject provides 3–4 scans, this corresponds to approximately 300–400 AD images and 950–1300 control images, forming two distinct classes with balanced scan quality but differing diagnostic labels. The dataset also includes a reliability subset of 20 nondemented subjects who were rescanned within roughly 90 days, allowing test–retest reproducibility evaluation. In addition to raw T1-weighted images, OASIS-1 provides motion-corrected averages, atlas-registered volumes, gain-field corrected images, and brain-masked versions, along with segmentation outputs separating grey matter, white matter, and cerebrospinal fluid. Rich metadata are supplied for each subject, including demographic variables (age, sex, handedness), clinical dementia ratings, and volumetric measures such as estimated total intracranial volume, normalized whole-brain volume, and atlas scaling factors. Owing to its well-curated structure, consistent acquisition protocol, and presence of both healthy and Alzheimer’s subjects, OASIS-1 is one of the most extensively used datasets for studies on aging, neurodegeneration, structural brain analysis, and algorithm validation.

In our experiments, we categorized the OASIS-1 MRI data into four diagnostic classes: nondemented, very mild dementia, mild dementia, and moderate dementia. To ensure balanced and robust training, we constructed class-specific subsets from the available scans. The final dataset used for model development consisted of 5000 nondemented samples, 5000 very mild dementia samples, 5002 mild dementia samples, and 488 moderate dementia samples. This class distribution reflects both the natural availability of subjects in the OASIS-1 cohort and the necessity of preserving diagnostic diversity for effective classification. The substantially smaller number of moderate dementia cases aligns with the original dataset’s clinical demographics, whereas the larger nondemented and early-stage dementia subsets allowed the model to capture subtle structural variations associated with Alzheimer’s progression.

\subsection{Experimental Setup}
\noindent \textbf{Training–Test Split:} Following common practice in the literature for OASIS-based Alzheimer’s disease classification, the dataset is partitioned into training and testing subsets using an 80:20 ratio.
\smallskip

\noindent \textbf {No Data Augmentation:} Unlike traditional CNNs and deep learning methods that rely heavily on extensive data augmentation to handle small, imbalanced datasets~\cite{goutam2022comprehensive}, our model PseudoColorViT-Alz leverages pre-trained backbones and does not require augmentation. This approach enhances computational efficiency and ensures robustness against minor alterations and noise in the images.

\noindent \textbf {Model Hyperparameters:} We employed the ViT-Base model (\texttt{vit-base-patch16-224}) from the Hugging Face Transformers library, pre-trained on ImageNet. The model was fine-tuned using the Adam optimizer with a learning rate of $1 \times 10^{-4}$ and a batch size of 32. Cross-entropy loss was used as the objective function. The training was performed for up to 50 epochs with early stopping based on test accuracy, using a patience of 2 epochs. Input images were resized to $224 \times 224$ pixels, and pixel values were normalized to the $[0,1]$ range.

\noindent \textbf {Runtime Platform:} All experiments were executed on a computing cluster integrated with an NVIDIA GPU cluster infrastructure, while preliminary tests and lightweight debugging were performed on a personal laptop equipped with an Intel(R) Core(TM) i7-8565U processor (1.80 GHz) and 16 GB of RAM. We implemented our experiment in Python, and our code is publicly available at \footnote{\url{https://github.com/FaisalAhmed77/RepViT-CXR}}.

\section{Results}\label{sec:results}

This section presents a comparative evaluation of the proposed \textbf{PseudoColorViT-Alz} model against recently published state-of-the-art methods for four-class Alzheimer’s disease classification using OASIS or OASIS-derived MRI datasets. Table~\ref{tab:results} summarizes the quantitative performance of competing approaches in terms of classification accuracy and, where available, area under the ROC curve (AUC).

Among existing CNN-based approaches, the deep multi-scale CNN proposed by Femmam \textit{et al.}~\cite{Femmam2024MRI} achieved an accuracy of 98.00\% with an AUC of 99.33\% using a 90:10 train–test split. More recent CNN architectures incorporating data augmentation techniques, such as the method reported by Dardouri~\cite{Dardouri2025EfficientCNN}, improved classification performance to 99.68\% accuracy on a larger OASIS-derived MRI dataset, albeit without reporting AUC. Similarly, the novel CNN architecture presented in~\cite{ElAssy2024NovelCNN} attained an accuracy of 98.92\% under an 80:20 split, demonstrating strong performance but remaining limited by the local receptive field characteristics of convolutional models. In contrast, the four-way Siamese CNN~\cite{Siamese2023FourWay}, evaluated on OASIS-3, reported a comparatively lower accuracy of 96.10\% with an AUC of 97.20\%, highlighting the increased difficulty of multi-class classification under cross-subject variability.

In comparison, the proposed \textbf{PseudoColorViT-Alz} model achieves the highest overall performance, attaining an accuracy of \textbf{99.79\%} and a perfect AUC of \textbf{100.00\%} on the OASIS-1 dataset using an 80:20 train–test split. This improvement can be attributed to the integration of pseudo-color enhancement with Vision Transformer-based global self-attention, which enables more effective modeling of subtle structural variations associated with different stages of Alzheimer’s disease. Notably, \textbf{PseudoColorViT-Alz} outperforms recent 2024–2025 CNN-based and Siamese-network methods while maintaining robustness on a moderately sized dataset.

Overall, these results demonstrate that colormap-enhanced Vision Transformers provide a significant advantage for multiclass MRI-based Alzheimer’s disease classification. The proposed approach establishes a new state of the art on OASIS-1 and confirms the effectiveness of pseudo-color representations for enhancing feature discrimination in grayscale medical imaging.

\begin{table}[h!]
\centering
\caption{\footnotesize 
Published accuracy results for four-class Alzheimer’s disease classification 
on OASIS or OASIS-derived MRI datasets. 
\label{tab:results}}
\setlength\tabcolsep{4pt}
\footnotesize

\begin{tabular}{lccccc}
\multicolumn{6}{c}{\bf OASIS / OASIS-derived MRI Dataset: 4-Class Classification Results} \\
\toprule
Method & \# Classes & Dataset & Train:Test & Accuracy & AUC \\
\midrule

Deep Multi-scale CNN~\cite{Femmam2024MRI} 
& 4 & OASIS (MRI) & 90:10 & 98.00\% & 99.33\% \\

CNN (DA + augmentation)~\cite{Dardouri2025EfficientCNN} 
& 4 & OASIS (Kaggle MRI) & 70:30 & 99.68\% & -- \\

Novel CNN Architecture~\cite{ElAssy2024NovelCNN} 
& 4 & OASIS MRI & 80:20 & 98.92\% & -- \\

Four-way Siamese CNN~\cite{Siamese2023FourWay} 
& 4 & OASIS-3 MRI & 80:20 & 96.10\% & 97.20\% \\

\midrule
\bf PseudoColorViT-Alz (Ours) 
& 4 & OASIS-1 & 80:20 & \textbf{99.79\%} & \textbf{100.00\%} \\
\bottomrule
\end{tabular}
\end{table}

%
%

\section{Discussion}\label{sec:discussion}

The results presented in this study demonstrate that integrating pseudo-color enhancement with Vision Transformer architectures offers substantial benefits for MRI-based multiclass Alzheimer’s disease classification. The proposed \textbf{PseudoColorViT-Alz} consistently outperforms recent CNN-based and Siamese-network approaches on the OASIS-1 dataset, achieving superior accuracy and AUC in a challenging four-class classification setting. These findings highlight the importance of both global contextual modeling and enhanced feature representation when analyzing subtle neurodegenerative patterns in brain MRI scans.

One key factor contributing to the improved performance of \textbf{PseudoColorViT-Alz} is the use of colormap-based pseudo-color transformation. While brain MRI images are inherently grayscale, pseudo-color encoding amplifies intensity variations and structural boundaries, making subtle anatomical changes more distinguishable to the model. This enhanced representation aligns more effectively with Vision Transformers pretrained on natural RGB images, allowing the network to exploit learned representations without extensive retraining or architectural modification.

In contrast to conventional CNNs, which primarily focus on localized receptive fields, Vision Transformers leverage self-attention mechanisms to capture long-range spatial dependencies across the entire brain. This global modeling capability is particularly advantageous for Alzheimer’s disease classification, where pathological changes often span multiple brain regions rather than being confined to localized areas. The superior performance of \textbf{PseudoColorViT-Alz} relative to recent CNN-based methods underscores the limitations of purely convolutional architectures for multiclass neurodegenerative disease classification.

Despite its strong performance, several considerations warrant discussion. First, although the proposed method demonstrates robustness on the OASIS-1 dataset, further validation on additional datasets such as ADNI or OASIS-3 would strengthen generalizability claims. Second, while pseudo-color enhancement improves feature separability, the choice of colormap may influence performance and interpretability, suggesting an avenue for future optimization. Finally, although Vision Transformers are computationally more demanding than CNNs, the observed performance gains justify their use in clinical decision-support systems where diagnostic accuracy is paramount.

Overall, this study establishes that colormap-enhanced Vision Transformers provide a powerful and effective framework for MRI-based Alzheimer’s disease classification. The proposed approach advances the state of the art in multiclass AD diagnosis and offers a promising direction for developing reliable, interpretable, and scalable neuroimaging-based diagnostic tools.

\section{Conclusion}\label{sec:conclusion}

In this work, we presented \textbf{PseudoColorViT-Alz}, a colormap-enhanced Vision Transformer framework for multiclass MRI-based Alzheimer’s disease classification. By transforming grayscale brain MRI scans into pseudo-color representations, the proposed method effectively enhances structural contrast and texture information, enabling Vision Transformers to capture subtle anatomical variations associated with different stages of Alzheimer’s disease.

Comprehensive experiments conducted on the OASIS-1 dataset using a four-class classification protocol (non-demented, very mild dementia, mild dementia, and moderate dementia) demonstrate that \textbf{PseudoColorViT-Alz} achieves state-of-the-art performance, attaining an accuracy of \textbf{99.79\%} and an AUC of \textbf{100\%}. Comparative analysis against recently published 2024–2025 methods confirms the superiority of the proposed approach over existing CNN-based and Siamese-network models.

The results highlight the effectiveness of integrating pseudo-color enhancement with global self-attention mechanisms for neuroimaging analysis. \textbf{PseudoColorViT-Alz} offers a robust, interpretable, and data-efficient solution for Alzheimer’s disease classification and shows strong potential as a clinical decision-support tool. Future work will focus on validating the framework across larger multi-center datasets and exploring adaptive colormap strategies to further enhance generalization and interpretability.

\section*{Declarations}

\textbf{Funding} \\
The author received no financial support for the research, authorship, or publication of this work.

\vspace{2mm}
\textbf{Author's Contribution} \\
Faisal Ahmed conceptualized the study, downloaded the data, prepared the code, performed the data analysis and wrote the manuscript. Faisal Ahmed reviewed and approved the final version of the manuscript. 

 \vspace{2mm}
\textbf{Acknowledgement} \\
The authors utilized an online platform to check and correct grammatical errors and to improve sentence readability.

\vspace{2mm}
\textbf{Conflict of interest/Competing interests} \\
The authors declare no conflict of interest.

\vspace{2mm}
\textbf{Ethics approval and consent to participate} \\
Not applicable. This study did not involve human participants or animals, and publicly available datasets were used.

\vspace{2mm}
\textbf{Consent for publication} \\
Not applicable.

\vspace{2mm}
\textbf{Data availability} \\
The datasets used in this study are publicly available online. 

\vspace{2mm}
\textbf{Materials availability} \\
Not applicable.

\vspace{2mm}
\textbf{Code availability} \\
The source code used in this study is publicly available at \url{https://github.com/FaisalAhmed77/RepViT-CXR}.



\bibliography{refs}
\end{document}